\documentclass[3p,times,two column]{elsarticle}

\usepackage{ecrc}

\volume{00}

\firstpage{1}

\journalname{Nuclear Physics B Proceedings Supplement}

\runauth{}

\jid{nuphbp}

\jnltitlelogo{Nuclear Physics B Proceedings Supplement}

\usepackage{amssymb}
\usepackage[figuresright]{rotating}

\begin{document}

\begin{frontmatter}



\dochead{}

\title{Upper Limit on the Diffuse $\nu_\mu$ Flux  with the ANTARES Telescope}


\author{Simone Biagi for the ANTARES collaboration}
\address{Dipartimento di Fisica dell'Universit\`a and INFN, 
		Viale Berti Pichat 6/2, 40127 Bologna, Italy}

\begin{abstract}
A search for very-high energy cosmic muon neutrinos from unresolved sources is presented using
data collected by the ANTARES neutrino  telescope. Data  corresponding to 334 days of equivalent live time  show that  the observed number of events is compatible with the expected number of background events.  A 90\% c.l. upper limit on the diffuse $\nu_\mu$ flux  is set at    
$E^2\Phi_{90\%}  =   5.3 \times 10^{-8}   \  \mathrm{GeV\ cm^{-2}\ s^{-1}\ sr^{-1}} $ 
in the energy range  20 TeV -- 2.5 PeV.
\end{abstract}


\end{frontmatter}

% \linenumbers

\begin{figure}[!b]
\vskip -.5cm
\center{\includegraphics[width=82mm]{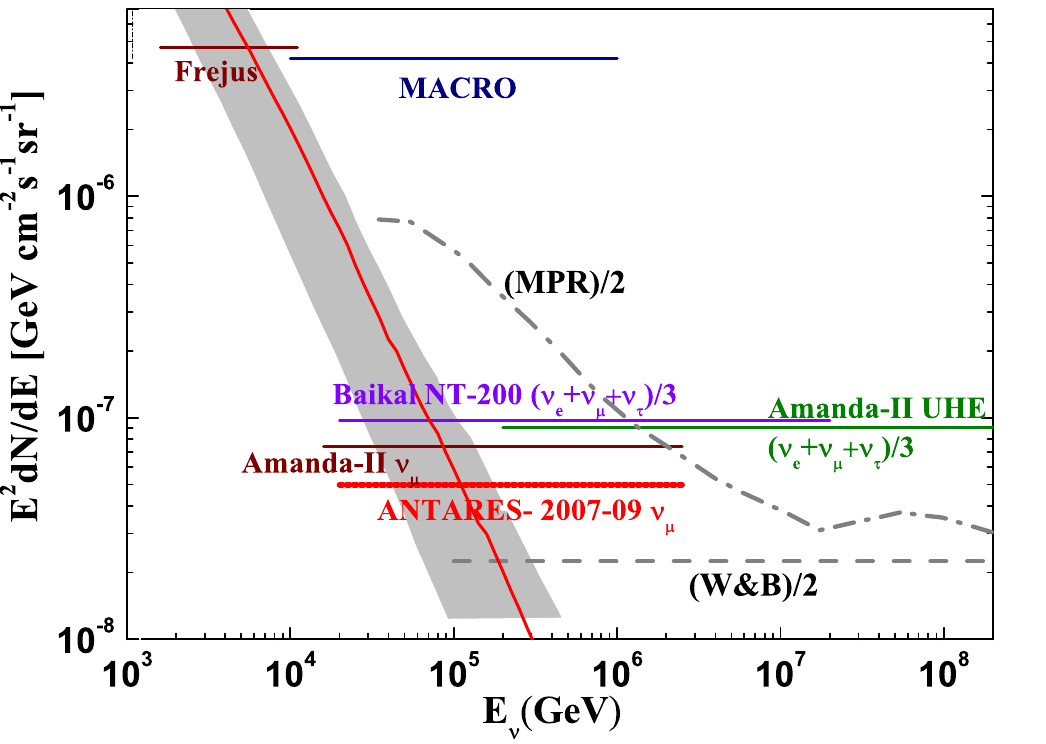}}
\vskip -.3cm
\caption{\small{
%%ANTARES 90\% c.l. upper limit  for a $E^{-2}$ diffuse  $\nu_\mu+\overline \nu_\mu$ flux, compared with the results obtained from  other experiments. 
%%The Frejus \cite{Frejus}, MACRO \cite{MACRO}, Amanda-II  \cite{Amanda_numu} limits refer to $\nu_\mu+\overline \nu_\mu$.
%%The Baikal \cite{Baikal} and Amanda-II  UHE  \cite{Amanda-UHE} refer to $\nu + \overline \nu$ of all-flavours, and are divided by 3. 
%%For reference, the W\&B \cite{wb} and the MPR  \cite{mpr} for transparent sources limits are also shown, divided by two to take into account neutrino oscillations.
The ANTARES 90\% c.l. upper limit  for a $E^{-2}$ diffuse  $\nu_\mu+\overline \nu_\mu$ flux, compared with the results obtained from  other experiments and theoretical predictions.
See \cite{chiarusi} and references therein.
}}
\label{fig:upper_limits}
\end{figure}

%\section{sezione 1}\label{}

The ANTARES  neutrino telescope is a three-dimensional array of 885 photomultiplier tubes (PMT) distributed over 12 lines installed in the Mediterranean Sea \cite{mar5line}. 
A search for a  diffuse flux of   muon neutrinos using data collected  from December 2007 to December 2009 is presented. 
%%The ANTARES  neutrino telescope is a three-dimensional array of 885 photomultipliers  (PMT)  installed in the Mediterranean Sea \cite{mar5line}. 
%%A search for a  diffuse flux of   muon neutrinos   is presented. 

Atmospheric muons  and neutrinos  are the main sources  of background in a neutrino telescope.  The former  can be suppressed by applying requirements on the  topology of the events, %\cite{aart_icrc09},  
the latter is an irreducible background. 
As the spectrum of cosmic neutrinos is expected to be harder than that of atmospheric neutrinos,  the signal searched for corresponds  to  an excess of high energy events,  produced  by astrophysical sources, in the 
measured energy spectrum without any particular assumption on the source  direction.  

A test signal spectrum  $\propto E_\nu^{-2}$ and the ``conventional'' atmospheric Bartol flux \cite{bartol} were simulated.
An   energy estimator \cite{biagi}, based on the mean number of hit repetitions ($R$)  on the PMTs, is used to separate the diffuse flux signal from the atmospheric    $\nu_\mu$ background.
A cut over the $R$ variable is optimized with the Model Rejection Procedure \cite{mrp} using Monte Carlo expectations  only. 

Nine high energy neutrino candidates are found with an expected background of $10.7\pm2$ events. The 90\% c.l. upper limit on the diffuse $\nu_\mu$ flux  including   systematic uncertainties is computed  with the method of \cite{conra}: it is  
\begin{equation}  
E^2\Phi_{90\%}  =   5.3 \times 10^{-8}   \  \mathrm{GeV\ cm^{-2}\ s^{-1}\ sr^{-1}} 
\end{equation} 
%\vskip -.1 cm
in the energy range  20 TeV -- 2.5 PeV. 
The result is compared  with other experiments  %and theoretical predictions  in Fig. \ref{fig:upper_limits}.
in Fig. \ref{fig:upper_limits}.

%% References


\bibliographystyle{elsarticle-num}
\bibliography{}



\end{document}